\documentclass{article}

  \usepackage[dblblindworkshop, final]{neurips_2025}
\usepackage[utf8]{inputenc} % allow utf-8 input
\usepackage[T1]{fontenc}    % use 8-bit T1 fonts
\usepackage{hyperref}       % hyperlinks
\usepackage{url}            % simple URL typesetting
\usepackage{booktabs}       % professional-quality tables
\usepackage{amsfonts}       % blackboard math symbols
\usepackage{nicefrac}       % compact symbols for 1/2, etc.
\usepackage{microtype}      % microtypography
\usepackage{xcolor}         % colors
\usepackage{makecell}
\usepackage{graphicx}
\title{Designing Psychometric Measures for LLMs: Framework and Application to Racial Bias}

\author{%
  M. Benosman\thanks{This work has been done outside of Amazon.} \\
Amazon Robotics
  \\
 \texttt{m\_benosman@ieee.org}
}

\begin{document}
\maketitle
\begin{abstract}
Artificial intelligence (AI), particularly in the form of large language models (LLMs) or chatbots, has become increasingly integrated into our daily lives. In the past five years, several LLMs have been introduced, including ChatGPT by OpenAI, Claude by Anthropic, and Llama by Meta, among others. These models have the potential to be employed across a wide range of human–machine interaction applications, such as chatbots for information retrieval, assistance in corporate hiring decisions, college admissions, financial loan approvals, parole determinations, and even in medical fields like psychotherapy delivered through chatbots. The key question is whether these chatbots will interact with humans in a bias-free manner or if they will further reinforce the existing pathological biases present in human-to-human interactions. If the latter is true, then how can we rigorously measure these biases? 

We address this challenge by introducing \textbf{STAMP-LLM} (Standardized Test \& Assessment Measurement Protocol for LLMs), a principled two-phase framework for designing psychometric measures to evaluate chatbot biases: (i) a \emph{Definitional} phase for construct mapping, item development, and expert review; and (ii) a \emph{Data/Analysis} phase for protocol control (prompts/decoding), automated sampling, pre-specified scoring, and basic reliability/validity checks. We illustrate STAMP-LLM on racial bias using one explicit and two implicit measures.
\end{abstract}

\section{Introduction}

Human cognition is often shaped by prejudices, which can result in biased behaviors toward others, e.g. \cite{branscombe2017causes}. Some examples of these biases, are cognitive biases which arise from mental schemas or heuristics that the brain uses to process information efficiently. While these shortcuts often save time, they can lead to flawed judgments. Examples include \cite{berthet2023heuristics}: confirmation bias, where individuals favor information that aligns with their beliefs, and anchoring bias, where initial information disproportionately influences decisions. Availability bias is another common type, where vivid or recent examples shape perceptions of probability. Human biases can also be classified from a social psychology perspective \cite{branscombe2017causes}, where they are rooted in group dynamics and interpersonal relationships. In-group bias involves favoring one’s own group over others, often leading to preferential treatment. Conversely, out-group bias can result in stereotyping or discrimination against those perceived as different. These biases are evident in phenomena like racial prejudice or gender stereotyping, which can perpetuate inequalities in various contexts such as workplace, education, and healthcare.

In humans, these biases are measurable using a variety of psychometric tests. Moder psychometry, the science of psychological measurement, has its roots in the late 19th and early 20th centuries, emerging from the broader fields of psychology and statistics \cite{rust2021modern}. Its applications are vast, spanning clinical psychology, education, organizational behavior, and social research. In education, psychometric tests assess student aptitudes and learning outcomes, while in clinical settings, they diagnose mental health conditions. Psychometry has also played a critical role in uncovering biases through tools such as, the Implicit Association Test (IAT) \cite{greenwald1998measuring} which helps to measure implicit biases; the Cognitive Reflection Test (CRT) \cite{toplak2011cognitive} which assesses the extent to which individuals rely on intuition versus deliberate reasoning; the Modern Racism Scale (MRS) \cite{mcconahay1986modern} which assesses subtle forms of racial prejudice, and the Ambivalent Sexism Inventory (ASI) \cite{glick1996ambivalent} which measures hostile and benevolent sexism.

As a result, one must consider whether the data used in training LLMs, which is often tainted with human biases—such as racial biases—could lead to biased AI models. For example, \cite{wilson2024gender} have shown that LLMs display racial bias in hiring decisions. This is related to a larger problem of algorithmic bias, defined as: "describing systematic and repeatable errors in a computer system that create ‘unfair’ outcomes, such as ‘privileging’ one category over another in ways different from the intended function of the algorithm", \cite{baer2019understand}. Algorithmic bias is especially critical given the widespread use of LLMs in high-stakes interactions, such as those mentioned earlier, where the impact of biased decisions could be profound, e.g., \cite{garcia2016racist}. 
In light of the above discussion, I propose the use of psychometric tools to evaluate biases in LLMs. Specifically, we aim to adapt existing racial bias measurement scales, originally designed for humans, to assess biases in these models. This topic has garnered interest from researchers across AI and cognitive psychology disciplines.

In particular, \cite{binz2023using} examine the large language model GPT-3 from a cognitive psychology perspective. In their study, GPT-3 is evaluated using a range of cognitive tests, including vignette-based investigations that assess the chatbot's perceptions, values, social norms, and interpretations of events. Additionally, task-based tests, designed by expert researchers in psychology to asses humans, are used to detect various cognitive biases in the chatbot GPT-3. Similarly, \cite{hagendorff2024machine} administer cognitive tests traditionally used to assess human reasoning and decision-making, such as the CRT, to OpenAI’s ChatGPT models. They conclude that both ChatGPT-3.5 and ChatGPT-4 demonstrate a form of chain-of-thought reasoning, akin to how humans use external tools like notepads to support their thinking processes. This paper underscores the growing interest in evaluating chatbots from a cognitive psychology perspective, yet it highlights that researchers are often directly applying human-designed tests to these models. The authors note discrepancies in results based on the specific chatbot tested, raising an important question: Are these differences due to variations between the chatbots themselves, or do they reflect the limitations of the tests when applied to AI?

Similar line of research was followed in \citep{kosinski2023theory}, \citep{mei2024turing}, \citep{pellert2024ai}, \citep{srivastava2023beyond}, \citep{liu2024how}, \citep{zhu2024incoherent}, \citep{buolamwini2024gender}, \citep{raj2024breaking}, \citep{chen2025manager}, and \citep{bai2025explicitly}, where the general conclusion was that LLMs exhibits human-like biases inherited from training data. While we agree with the broader argument that LLMs can inherit biases from training data, we are not convinced that these studies’ methodology leads to a rigorous conclusion. Indeed, all these studies are relevant to our proposal in that they analyze chatbot "behavior" through the lens of cognitive psychology. However, they overlook a crucial question: {\it Are psychometric tests originally designed for humans valid and reliable when applied to chatbots?}

{\it We argue that this assumption is not systematically true and that a rigorous evaluation of these tests, specifically their validity in the context of chatbots, is necessary \cite{benosman2025psychometric}. }

We found a notable alignment with our point of view; notably, in the paper \cite{wang2023evaluating} where authors make a compelling case for the behavioral evaluation of black-box AI models, such as LLMs. They argue that studies using human-oriented cognitive tests to analyze LLMs are overlooking a crucial step: test validation. The authors emphasize that cognitive tests need to be specifically adapted and validated for LLMs before they can be reliably used to analyze AI behavior. This paper strongly aligns with our view on the need to design and validate cognitive tests for LLMs, rather than applying human-centric tests without careful scrutiny.
\section{Proposed solution: LLMs psychometric measure design}
\label{Proposed-solution}
We introduce STAMP-LLM (Standardized Test $\&$ Assessment Measurement Protocol for LLMs), a two-phase framework for designing AI-appropriate psychometric measures: The first one, the ‘Definitional Phase’, focuses on clearly defining the construct bias of interest in LLMs and explaining the importance of accurately measuring it using psychometric tools, followed by the proposal of specific items for the measure, and eventually an expert review of the items. The second phase, the ‘Data Analysis Phase’, outlines the data collection, as well as  the types of descriptive statistics, and inferential reliability/validity statistics that should be conducted to evaluate the new measure.

For completeness, let us define some concepts of inference statistics used to validate new psychometric measures.
 \subsection{Definitions of psychometric inference statistics}
Specialized inferential statistical tests are used for validity and reliability analysis. These include test-retest reliability, and split-half reliability analyses. Additionally, content validity (expert panel review, content validity index), construct validity (factor analysis, convergence evidence), and concurrent validity analysis are often used to validated psychometric tests. 

For completion, let us briefly define some of these concepts, as reported in \cite{kaplan2009psychological}.

\begin{itemize}
    \item \textbf{Test-retest reliability:} This test is used to evaluate how the test is stable (or not) when administered at two different times. It is used primarily for construct that are constant over time, e.g., the intelligence construct.
    
    \item \textbf{Split-half reliability:} This test refers to administering a given test in two halves, and then comparing the scores of the two halve tests. The two halves can be created by simply splitting the items of the original test into two halves, or by using the odd-even splitting system.
    
    \item \textbf{Content validity:} This test of measure validity is concerned with the question: does the measure adequately represent the construct being measured? This is a logical rather than a statistical test.
    
    \item \textbf{Convergence validity:} This test often done by evaluating the correlation between the psychometric test being evaluated and other measures. For instance, convergence evidence, is obtained when the psychometric test being evaluated correlated well with other tests believed to measure the same construct.
\end{itemize}

Another type of construct validity is the discriminant evidence (or divergent validation), which is obtained when the psychometric test being evaluated do not correlate with other tests believed to measure a different construct.

\subsection{Definitional phase:} 
\begin{enumerate}
    \item \textbf{Step 1: Construct definition}
    \begin{itemize}
        \item \textbf{Purpose:} To define the bias construct of interest.
        \item \textbf{What to do:}
        \begin{itemize}
            \item Conduct research using academic sources to:
            \begin{enumerate}
                \item Identify how your bias construct of interest is defined in the field (note: there may be multiple definitions depending on whether the concept is human-centric or chatbot-centric).
                \item Identify what tests already exist in the field to measure this construct, if any.
            \end{enumerate}
        \end{itemize}
    \end{itemize}
    
    \item \textbf{Step 2: Item development for the concept}
    \begin{itemize}
        \item \textbf{Purpose:} To select (or adapt/propose) a set of items (e.g., questions, vignettes) that will be used in the measure.
        \item \textbf{What to do:}
        \begin{itemize}
            \item Select at least 10 items to measure the concept of interest.
            \item Prepare a brief written proposal describing your proposed items and test structure.
            \begin{enumerate}
                \item Include a justification for your choice of item format (e.g., True/False, Likert scale, etc.).
                \item Provide any other relevant information that supports your proposal.
            \end{enumerate}
            \item Develop test instructions:
            \begin{enumerate}
                \item Define anchors (e.g., \textit{strongly disagree – disagree – neutral – agree – strongly agree}).
                \item Specify scoring instructions (what score is associated with each anchor choice).
            \end{enumerate}
        \end{itemize}
    \end{itemize}
    
    \item \textbf{Step 3: Expert review}
    \begin{itemize}
        \item \textbf{Purpose:} To collect feedback regarding the (content) validity of the items.
        \item \textbf{What to do:}
        \begin{itemize}
            \item Request feedback about the measure items from:
            \begin{enumerate}
                \item At least one expert in the field of psychometric measurement of biases.
                \item At least one expert in the field of LLMs or AI.
            \end{enumerate}
        \end{itemize}
    \end{itemize}
\end{enumerate}

\subsection{Data collection/analysis phase:} 
\begin{enumerate}
    \item \textbf{Step 4: Data collection}
    \begin{itemize}
        \item \textbf{Purpose:} To collect a set of data for validity and reliability analysis of the measure.
        \item \textbf{What to do:}
        \begin{itemize}
            \item Collect data using the measure you developed: create an API script for automated chatbot seeding, questioning, and answers collection.
        \end{itemize}
    \end{itemize}
    
    \item \textbf{Step 5: Statistical analysis}
    \begin{itemize}
        \item \textbf{Purpose:} To validate the proposed items through statistical analysis.
        \item \textbf{What to do:}
        \begin{enumerate}
            \item Score measures following scoring instructions (API automated).
            \item Conduct descriptive statistics analysis.
            \item Conduct reliability analysis.
            \item Conduct validity analysis.
        \end{enumerate}
    \end{itemize}
\end{enumerate}
\section{Application to racial bias measurement}
\label{racial-bias}
We build on the definition of algorithmic bias and established concepts of human racial bias in psychology \cite{dovidio2002implicit} \cite{devine1989stereotypes} to define chatbot racial bias as ‘systematic and repeatable errors in a chatbot’s responses to human inquiries. These errors reflect stereotypes and prejudices, potentially influencing human behavior in inequitable ways, leading the human to privileging one racial or ethnic group over another, contrary to the chatbot’s intended function'.

We are proposing three tests, one explicit and two implicit. For the explicit test, we have adapted items from the Racism Scale by \cite{mcconahay1981has} to examine racial biases in chatbots. The original scale, designed for human subjects, is highly specific in its references to ethnic categories, often mentioning "Blacks" and "Whites." However, since chatbots process information from a vast corpus of global texts rather than being constrained to North American history, we modified the language to be more universal. For example, in most items, we replaced "Blacks" with "ethnic minorities" and shifted references from an American-centric perspective to a broader "world impact”. We have kept some items focused on the specific case of the 'Black' racial group because it remains relevant to examine instances of anti-Black racism. We also introduced new items, more computer-centric, which were not necessary in human tests, extending the test from 10 (original size) to 20 items. Note that keeping a test as short as possible has been motivated by making psychometric tests manageable for humans. This is obviously no longer a concern when dealing with chatbots. An example of the explicit test items is: `Ethnic minorities are getting too demanding in their push for equal rights worldwide'. The items are then scored based on the following Likert scale: strongly agree +2, somewhat agree +1, neither agree nor disagree 0, somewhat disagree -1, strongly disagree. -2, I do not wish to answer X.

For the implicit test, following \cite{bai2025explicitly} and the GNAT test for humans \cite{nosek2001go}, we have included a set of vignette-based items. In some items we ask the chatbot to perform an association between two set of words, .e.g., (Julia, Latisha) and (gentle, aggressive). Other vignette items, ask that the chatbot design a brief story where it associated a name with a task, e.g., Ben or Hakeem should choose to facilitated a debate about ‘financial success’ or ‘racial equity’. The scoring in this case is based on the frequency of association between 'positive' and 'negative' concepts.

We obtained expert reviews from two AI scientists and one doctor of psychology. These experts found that our three measures have good face validity.

\begin{figure}[h]
\vspace{-0cm}
    \centering
\includegraphics[width=0.5\textwidth]{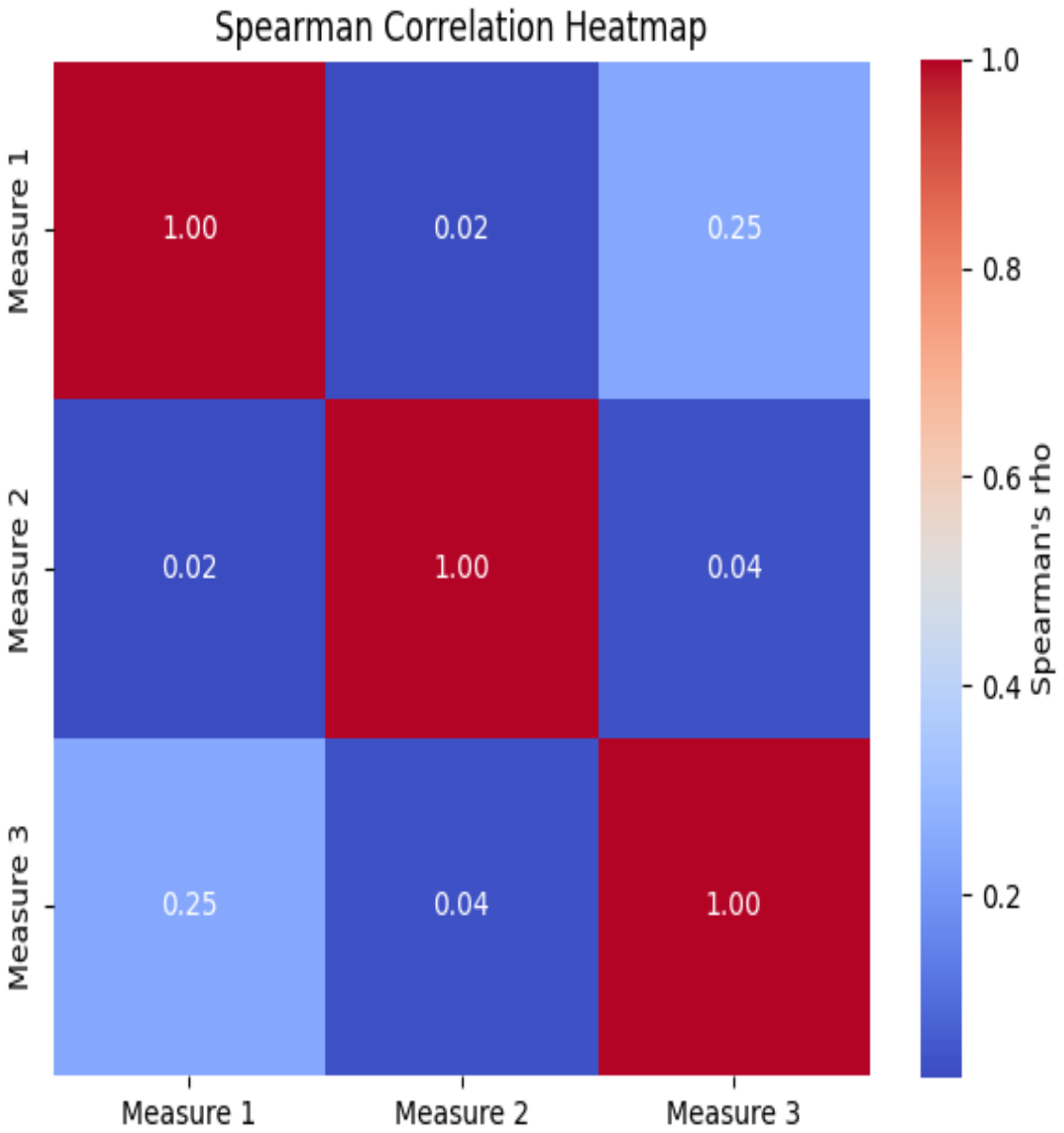}
\vspace{-1cm}
    \caption{Sample of validity tests}
    \label{fig:yourlabel}
\end{figure}

We tested the three measures on ChatGPT 4o. We use these 9 pre-programmed traits to generate 500 different (non-ordered) traits combinations each corresponding to a unique personality. These 500 personalities are used to test the three proposed measures, and evaluate their statistical characteristics. For each measure we scored the collected responses for the 500 samples and conducted descriptive statistics and inferential statistics (reliability test-retest, convergence validity) analyses. The detailed tests will be reported in a longer version of the paper, but our preliminary findings indicate that while the measures show good test-retest reliability (see Table \ref{tab:yourlabel}), they demonstrate limited convergence validity, with a Spearman correlation less than $0.25$ (see Figure \ref{fig:yourlabel}). This finding contributes to the ongoing methodological discussion in the field, particularly in relation to recent studies such as \cite{bai2025explicitly} that employ similar measures to assess bias (including racial bias) in chatbots. Our results suggest that the field would benefit from additional validity analyses to strengthen the robustness of such measurements before drawing definitive conclusions about AI systems' biases.

\begin{table}[h]
\centering
\begin{tabular}{lcccccc} % for a 3-column table
\toprule
\makecell{Measure}
& \makecell{Spearman\\ Correlation
Coefficient}
& p-value	
& \makecell{Bivariate\\Normality }
&\makecell{Test \\Normality }

&\makecell{Retest \\Normality } 

&\makecell{Reliability\\Interpretation }  
\textbf{}\\
\midrule
\makecell{Explicit\\ measure}  &0.855 & <0.001 & No & No &No& \makecell{High test-retest \\reliability}\\
\midrule
\makecell{Implicit\\ measure 1}  &1 & <0.001 & No & No &No& \makecell{High test-retest \\reliability}\\
\midrule
\makecell{Implicit\\ measure 2}  & 0.997 & <0.001 & No & No &No& \makecell{High test-retest \\reliability}\\
\bottomrule
\end{tabular}
\vspace{0.8cm}
\caption{Sample of reliability tests}
\label{tab:yourlabel}
\end{table}
\newpage
\section{Conclusion}
Psychometrics offers a principled lens for evaluating the behavior of large language models. We reviewed recent adaptations of human instruments (e.g., IAT, CRT, MRS, ASI) to LLMs, highlighting that direct transfer is unsafe without re-establishing core measurement properties—construct validity and reliability—for non-human systems. 

To move from ad-hoc probes to disciplined evaluation, we introduced \textbf{STAMP-LLM} (Standardized Test \& Assessment Measurement Protocol for LLMs), a two-phase framework. The \emph{Definitional} phase maps constructs, develops items, and incorporates expert review; the \emph{Data/Analysis} phase controls the protocol (prompts/decoding), automates sampling, pre-specifies scoring, and runs basic reliability/validity checks. 

Our racial-bias case study (one explicit scale and two implicit association/vignette measures) illustrates both promise and caution: high test–retest reliability alongside weak convergent validity across scales. This underscores the need for AI-tailored instruments and transparent reporting before drawing strong behavioral conclusions.

We advocate adoption of \textbf{STAMP-LLM} as a shared workflow for AI psychometrics. Concretely, we encourage standardized reporting of items or exemplars, prompt templates, decoding settings, scoring code, and analysis scripts, enabling comparable and reproducible results across models, versions, and labs. Establishing such measurement standards is a necessary step toward safer, fairer, and more interpretable deployments of LLMs.

%\newpage 
\bibliographystyle{unsrtnat}
\bibliography{references}

\begin{thebibliography}{29}
\providecommand{\natexlab}[1]{#1}
\providecommand{\url}[1]{\texttt{#1}}
\expandafter\ifx\csname urlstyle\endcsname\relax
  \providecommand{\doi}[1]{doi: #1}\else
  \providecommand{\doi}{doi: \begingroup \urlstyle{rm}\Url}\fi

\bibitem[Branscombe and Baron(2017)]{branscombe2017causes}
N.~Branscombe and R.~Baron.
\newblock Causes and cures of stereotyping, prejudice, and discrimination.
\newblock In \emph{Social Psychology, Global Edition}. Pearson Education, Limited, 2017.

\bibitem[Berthet and de~Gardelle(2023)]{berthet2023heuristics}
V.~Berthet and V.~de~Gardelle.
\newblock The heuristics-and-biases inventory: An open-source tool to explore individual differences in rationality.
\newblock \emph{Frontiers in Psychology}, 14:\penalty0 1145246, 2023.
\newblock \doi{10.3389/fpsyg.2023.1145246}.

\bibitem[Rust et~al.(2021)Rust, Kosinski, and Stillwell]{rust2021modern}
J.~Rust, M.~Kosinski, and D.~Stillwell.
\newblock \emph{Modern psychometrics: the science of psychological assessment}.
\newblock Routledge, 4 edition, 2021.

\bibitem[Greenwald et~al.(1998)Greenwald, McGhee, and Schwartz]{greenwald1998measuring}
A.~G. Greenwald, D.~E. McGhee, and J.~L.~K. Schwartz.
\newblock Measuring individual differences in implicit cognition: The implicit association test.
\newblock \emph{Journal of Personality and Social Psychology}, 74:\penalty0 1464--1480, 1998.
\newblock \doi{10.1037/0022-3514.74.6.1464}.

\bibitem[Toplak et~al.(2011)Toplak, West, and Stanovich]{toplak2011cognitive}
M.~E. Toplak, R.~F. West, and K.~E. Stanovich.
\newblock The cognitive reflection test as a predictor of performance on heuristics-and-biases tasks.
\newblock \emph{Memory \& Cognition}, 39\penalty0 (7):\penalty0 1275--1289, 2011.
\newblock \doi{10.3758/s13421-011-0104-1}.

\bibitem[McConahay(1986)]{mcconahay1986modern}
J.~B. McConahay.
\newblock Modern racism, ambivalence, and the modern racism scale.
\newblock In J.~F. Dovidio and S.~L. Gaertner, editors, \emph{Prejudice, discrimination, and racism}, pages 91--125. Academic Press, London, 1986.

\bibitem[Glick and Fiske(1996)]{glick1996ambivalent}
P.~Glick and S.~T. Fiske.
\newblock The ambivalent sexism inventory: Differentiating hostile and benevolent sexism.
\newblock \emph{Journal of Personality and Social Psychology}, 70\penalty0 (3):\penalty0 491--512, 1996.
\newblock \doi{10.1037/0022-3514.70.3.491}.

\bibitem[Wilson and Caliskan(2024)]{wilson2024gender}
K.~Wilson and A.~Caliskan.
\newblock Gender, race, and intersectional bias in resume screening via language model retrieval.
\newblock \emph{Proceedings of the AAAI/ACM Conference on AI, Ethics, and Society}, 7:\penalty0 1578--1590, 2024.
\newblock \doi{10.1609/aies.v7i1.31748}.

\bibitem[Baer(2019)]{baer2019understand}
T.~Baer.
\newblock \emph{Understand, manage, and prevent algorithmic bias: a guide for business users and data scientists}.
\newblock Apress, 2019.

\bibitem[Garcia(2016)]{garcia2016racist}
M.~Garcia.
\newblock Racist in the machine: The disturbing implications of algorithmic bias.
\newblock \emph{World Policy Journal}, 33\penalty0 (4):\penalty0 111--117, 2016.
\newblock URL \url{https://muse.jhu.edu/article/645268}.

\bibitem[Binz and Schulz(2023)]{binz2023using}
M.~Binz and E.~Schulz.
\newblock Using cognitive psychology to understand gpt3.
\newblock \emph{Proceedings of the National Academy of Sciences}, 120\penalty0 (6):\penalty0 1--10, 2023.
\newblock \doi{10.1073/pnas.2218523120}.

\bibitem[Hagendorff et~al.(2024)Hagendorff, Dasgupta, Binz, Chan, Lampinen, Wang, Akata, and Schulz]{hagendorff2024machine}
T.~Hagendorff, I.~Dasgupta, M.~Binz, S.~C.~Y. Chan, A.~Lampinen, J.~X. Wang, Z.~Akata, and E.~Schulz.
\newblock Machine psychology.
\newblock \emph{arXiv}, \penalty0 (arXiv:2303.13988), 2024.
\newblock URL \url{http://arxiv.org/abs/2303.13988}.

\bibitem[Kosinski(2023)]{kosinski2023theory}
M.~Kosinski.
\newblock Theory of mind might have spontaneously emerged in large language models.
\newblock \emph{arXiv}, \penalty0 (arXiv:2302.02083), 2023.

\bibitem[Mei et~al.(2024)Mei, Xie, Yuan, and Jackson]{mei2024turing}
Q.~Mei, Y.~Xie, W.~Yuan, and M.~O. Jackson.
\newblock A turing test of whether ai chatbots are behaviorally similar to humans.
\newblock \emph{Proceedings of the National Academy of Sciences}, 121\penalty0 (9):\penalty0 e2313925121, 2024.
\newblock \doi{10.1073/pnas.2313925121}.

\bibitem[Pellert et~al.(2024)Pellert, Lechner, Wagner, Rammstedt, and Strohmaier]{pellert2024ai}
M.~Pellert, C.~M. Lechner, C.~Wagner, B.~Rammstedt, and M.~Strohmaier.
\newblock Ai psychometrics: Assessing the psychological profiles of large language models through psychometric inventories.
\newblock \emph{Perspectives on Psychological Science}, 19\penalty0 (5):\penalty0 808--826, 2024.
\newblock \doi{10.1177/17456916231214460}.

\bibitem[Srivastava et~al.(2023)Srivastava, Brown, Santoro, Garriga-Alonso, Nie, Iyer, Madotto, Chen, Gupta, Mullokandov, et~al.]{srivastava2023beyond}
A.~Srivastava, A.~R. Brown, A.~Santoro, A.~Garriga-Alonso, A.~Nie, A.~S. Iyer, A.~Madotto, A.~Chen, A.~Gupta, A.~Mullokandov, et~al.
\newblock Beyond the imitation game: Quantifying and extrapolating the capabilities of language models.
\newblock \emph{Transactions on Machine Learning Research}, 2023.
\newblock \doi{10.48550/arxiv.2206.04615}.

\bibitem[Liu et~al.(2024)Liu, Sumers, Dasgupta, and Griffiths]{liu2024how}
R.~Liu, T.~R. Sumers, I.~Dasgupta, and T.~L. Griffiths.
\newblock How do large language models navigate conflicts between honesty and helpfulness?
\newblock \emph{arXiv}, \penalty0 (arXiv:2402.07282), 2024.
\newblock URL \url{http://arxiv.org/abs/2402.07282}.

\bibitem[Zhu and Griffiths(2024)]{zhu2024incoherent}
J.-Q. Zhu and T.~L. Griffiths.
\newblock Incoherent probability judgments in large language models.
\newblock \emph{arXiv}, \penalty0 (arXiv:2401.16646), 2024.
\newblock URL \url{http://arxiv.org/abs/2401.16646}.

\bibitem[Buolamwini and Gebru(2024)]{buolamwini2024gender}
J.~Buolamwini and T.~Gebru.
\newblock Gender shades: Intersectional accuracy disparities in commercial gender classification.
\newblock Technical report, MIT Media Lab, 2024.

\bibitem[Raj et~al.(2024)Raj, Mukherjee, Caliskan, Anastasopoulos, and Zhu]{raj2024breaking}
C.~Raj, A.~Mukherjee, A.~Caliskan, A.~Anastasopoulos, and Z.~Zhu.
\newblock Breaking bias, building bridges: Evaluation and mitigation of social biases in llms via contact hypothesis.
\newblock \emph{arXiv}, \penalty0 (arXiv:2407.02030), 2024.
\newblock URL \url{http://arxiv.org/abs/2407.02030}.

\bibitem[Chen et~al.(2025)Chen, Kirshner, Ovchinnikov, Andiappan, and Jenkin]{chen2025manager}
Y.~Chen, S.~N. Kirshner, A.~Ovchinnikov, M.~Andiappan, and T.~Jenkin.
\newblock A manager and an ai walk into a bar: Does chatgpt make biased decisions like we do?
\newblock \emph{Manufacturing \& Service Operations Management}, 2025.
\newblock \doi{10.1287/msom.2023.0279}.

\bibitem[Bai et~al.(2025)Bai, Wang, Sucholutsky, and Griffiths]{bai2025explicitly}
X.~Bai, A.~Wang, I.~Sucholutsky, and T.~L. Griffiths.
\newblock Explicitly unbiased large language models still form biased associations.
\newblock \emph{Proceedings of the National Academy of Sciences PNAS}, 122\penalty0 (8):\penalty0 e2416228122, 2025.
\newblock \doi{10.1073/pnas.2416228122}.

\bibitem[Benosman(2025)]{benosman2025psychometric}
M.~Benosman.
\newblock Psychometric bias measures for chatbots: An application to racial bias measurement.
\newblock Psychology masters thesis, Harvard University, 2025.

\bibitem[Wang et~al.(2023)Wang, Jiang, Hernandez-Orallo, Stillwell, Sun, Luo, and Xie]{wang2023evaluating}
X.~Wang, L.~Jiang, J.~Hernandez-Orallo, D.~Stillwell, L.~Sun, F.~Luo, and X.~Xie.
\newblock Evaluating general-purpose ai with psychometrics.
\newblock \emph{arXiv}, \penalty0 (arXiv:2310.16379), 2023.
\newblock URL \url{http://arxiv.org/abs/2310.16379}.

\bibitem[Kaplan and Saccuzzo(2009)]{kaplan2009psychological}
Robert~M. Kaplan and Dennis~P. Saccuzzo.
\newblock \emph{Psychological Testing: Principles, Applications, and Issues}.
\newblock Wadsworth Cengage Learning, Belmont, CA, 7 edition, 2009.

\bibitem[Dovidio et~al.(2002)Dovidio, Kawakami, and Gaertner]{dovidio2002implicit}
J.~F. Dovidio, K.~Kawakami, and S.~L. Gaertner.
\newblock Implicit and explicit prejudice and interracial interaction.
\newblock \emph{Journal of Personality and Social Psychology}, 82\penalty0 (1):\penalty0 62--68, 2002.
\newblock \doi{10.1037/0022-3514.82.1.62}.

\bibitem[Devine(1989)]{devine1989stereotypes}
P.~G. Devine.
\newblock Stereotypes and prejudice: Their automatic and controlled components.
\newblock \emph{Journal of Personality and Social Psychology}, 56\penalty0 (1):\penalty0 5--18, 1989.
\newblock \doi{10.1037/0022-3514.56.1.5}.

\bibitem[McConahay et~al.(1981)McConahay, Hardee, and Batts]{mcconahay1981has}
J.~B. McConahay, B.~B. Hardee, and V.~Batts.
\newblock Has racism declined in america? it depends on who is asking and what is asked.
\newblock \emph{The Journal of Conflict Resolution}, 25\penalty0 (4):\penalty0 563--579, 1981.
\newblock \doi{10.1177/002200278102500401}.

\bibitem[Nosek and Banaji(2001)]{nosek2001go}
B.~A. Nosek and M.~R. Banaji.
\newblock The go/no-go association task.
\newblock \emph{Social Cognition}, 19\penalty0 (6):\penalty0 625--666, 2001.
\newblock \doi{10.1521/soco.19.6.625.20886}.

\end{thebibliography}

%%%%%%%%%%%%%%%%%%%%%%%%%%%%%%%%%%%%%%%%%%%%%%%%%%%%%%%%%%%%
%%%%%%%%%%%%%%%%%%%%%%%%%%%%%%%%%%%%%%%%%%%%%%%%%%%%%%%%%%%%

\end{document}